\documentclass{article}

\usepackage{arxiv}

\usepackage[utf8]{inputenc} 
\usepackage[T1]{fontenc}    
\usepackage{hyperref}       
\usepackage{url}            
\usepackage{booktabs}       
\usepackage{amsfonts}       
\usepackage{nicefrac}       
\usepackage{microtype}      
\usepackage{lipsum}		
\usepackage{graphicx}
\usepackage{natbib}
\usepackage{comment}
\usepackage{doi}

\title{A Survey for Real-Time Network Performance Measurement via Machine Learning}

\date{December 25, 2020} 	

\author{ \href{https://orcid.org/0000-0001-5257-7918}{\includegraphics[scale=0.06]{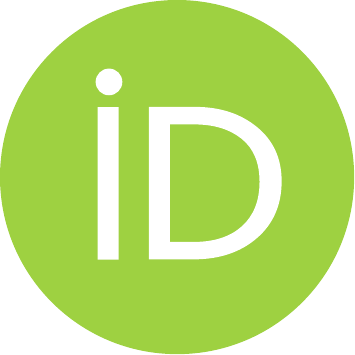}\hspace{1mm}Chien-Cheng Wu}\\
	\texttt{ccw@ieee.org} \\
}

\renewcommand{\headeright}{}
\renewcommand{\undertitle}{}

\hypersetup{
pdftitle={A Survey for Real-Time Network Performance Measurement via Machine Learning},
pdfsubject={q-bio.NC, q-bio.QM},
pdfauthor={Chien-Cheng WU},
pdfkeywords={Real-Time Network, Time-Senstive Network, Network Calculus, Machine Learning, Performance},
}

\begin{document}
\maketitle

\begin{abstract}
	Real-Time Networks (RTNs) provide latency guarantees for time-critical applications and it aims to support different traffic categories via various scheduling mechanisms. 
	Those scheduling mechanisms rely on a precise network performance measurement to dynamically adjust the scheduling strategies. 
	Machine Learning (ML) offers an iterative procedure to measure network performance. Network Calculus (NC) can calculate the bounds for the main performance indexes such as latencies and throughputs in an RTN for ML. 
	Thus, the ML and NC integration improve overall calculation efficiency. 
	This paper will provide a survey for different approaches of Real-Time Network performance measurement via NC as well as ML and present their results, dependencies, and application scenarios.
\end{abstract}


\section{Introduction}
\label{sec:introduction}

\subsection{Motivation}
Many scheduling algorithms in Real-Time Network rely on accurate performance estimations to adjust their operating strategies. 
The static performance estimations cannot adapt to the dynamic network changes and learn from the network feedback. 
The machine learning based approaches can eliminate these constraints via historical training data. 
The latency, throughput and reliability are the main performance indexes to be investigated. Network Calculus provides a formal framework to estimate the performance indexes. 
Thus, leveraging machine learning and Network Calculus can improve the Real-Time network performance measurement.

\subsection{Contribution of This Survey Article}
In this survey article, we make the following contributions:
\begin{itemize}
	\item We survey and classify advantages of ML-based network performance measurement in Real-Time Network.
	\item We discuss ML-based Real-Time Network architectures.
	\item We provide an in-depth discussion on ML strategies for network performance measurement.
	\item We outline open issues, challenges, and future research directions related to ML-based network performance measurement.
\end{itemize}

\subsection{Article Structure}
The paper is organized as follows: Comparison with related survey articles is presented in Section \ref{sec:latency}, Section \ref{sec:reliability} and Section \ref{sec:throughput}.  In the same sections, we also highlight the architectures for ML-based network performance measurement and ML strategies of network performance measurement. Moreover, case studies on the use of ML in network performance measurement are also presented in those sections. Issues, challenges, and future research directions are mentioned in Section \ref{sec:challenges}. Finally, Section \ref{sec:conclusion} concludes the paper.

\section{Latency}
\label{sec:latency}
Many RTN applications require deterministic bounds on the end-to-end delay. 
Network Calculus (NC) is an iterative analysis framework for the derivation of delay bounds. 
The models and analysis tools from NC generate all steps towards the derivation of delay bounds. 
However, \cite{KGS10-1} showed that this method suffers from the vast computational effort. 
The cost of delay analysis increases fast with the size and complexity of a network. Neural networks for graphs have recently been introduced to map network topologies and flows to graphs. 
This approach has been used in a variety of domains such as performance evaluation of networks with TCP flows \cite{10.1145/3150928.3150941}, routing protocols \cite{10.1145/3229607.3229610}, or basic logical reasoning tasks and program verification \cite{li2016gated}.

One example in \cite{8493033}, each server is represented as a node in the graph, with edges corresponding to the connections between servers. 
Each flow is represented as a node with edges connecting it to the path of traversed servers. 
The method to transferring those graph inputs for a neural network was able to process any general graphs. The authors demonstrated this method via a numerical evaluation and showed that it can be used at a small computational cost compared to traditional network analyzes. 
The other example uses machine learning to estimate service curves from measurements \cite{9219693}. Through service curves, the correct bounds on the worst-case flow delays and inferences cannot be computed precisely due to uncontrollable uncertainties introduced by measurements. But the authors implemented an iterative method to adaptively change the probe rate and improve accuracy by reducing bias and variability.

\section{Reliability}
\label{sec:reliability}
Nowadays, mission-critical applications had been widely discussed over the world. 
These applications rely on RTN to provide a higher level of Quality of Service (QoS). 
Reliability (loss) is one of the significant parameters to estimate the QoS within the RTN.
For instance, the network traffic management algorithms (such as connection control, flow control) require the loss analysis to make its scheduling decision \cite{8187644}.
Due to larger and more complex RTN development, the size and complexity of the loss estimation grow rapidly.
On the other hand, the difficulty in applying NC in realistic network scenarios is that links (or servers) may be unreliable and some packets are lost. 
Some NC loss analyses have proposed in \cite{1633336} and \cite{5471427} seek to estimate the packet loss from expectations instead of probabilities.
But the above methods cannot adapt to the scalable network changes.
One earliest work was introduced in  \cite{6662955}. The authors modeled unreliable networks using the stochastic NC and integrated the model with a retransmission-based loss recovery. From their numerical experiment, a small number of retransmission attempts already lends RTN to a delay bound’s blow-up.
Another related work in \cite{8514938} investigates reliability and NC performance within an RTN. The authors figure out that the RTN suffers from reliability and reproducibility issues during the NC computation and improved the NC performance by parallelizing its computation procedure.
However, to the best of the authors' knowledge, the state-of-the-art research on reliability investigation only via NC.
Therefore, machine learning and NC become a new subject of interest to investigate the tradeoff between accuracy and scalability of a loss analysis.

\section{Throughput}
\label{sec:throughput}
Under given delay constraints, the traffic carrying capacity (throughput) of RTN is another fundamental index for network management. \cite{1611082} \cite{5415864} NC is a more general theory that has been applied to predict the traffic carrying capacity of RTN. The goal of throughput estimation is to infer the available throughput of a network path using only external observations of data packets. 
To model the minimal available throughput, we should find the tightest link which has the smallest capacity.
In addition, the end-to-end available throughput of a network path is determined by the tightest link in the path.
If the rate of the cross-traffic dynamically changes during an estimation, or some packets are randomly lost and then retransmitting, the corresponding estimates are imprecise.

In \cite{KHANGURA201918}, \cite{10.1007/978-3-319-30505-9_30} the authors train a neural network using vectors constructed by packets.
The vectors contain the available bandwidth of the packet dispersion.
The neural network can generalize non-locally which kernel or ensemble machines with standard generic kernels are not able to do. It can recognize complicated functions even in the presence of noise and variability.



\section{Challenges and Future Research Directions}
\label{sec:challenges}
Aiming at an optimal balance between the feasible resource allocations and offloading, the network operators must have a deep understanding of network conditions. This not only calls for accurate models but also further raises the computational complexity.

Future research should generate much interest and progress with respect to ML extensions for reliability and throughput measurements. On the other hand, NC tools development is also one potential direction for next-generation RTNs. Tool support for network calculus has not been addressed by the prior-art and brings about a new interesting perspective that can accommodate the RTN loss during the measurement.

\section{Conclusion}
\label{sec:conclusion}
The real-time network is required for next-generation communication. 
To achieve the stringent goals of  real-time network, network operators rely on an efficient, reliable, flexible, and globally network performance measurement, which helps to assist real-time networks in providing these services promptly. 
The complex measurement becomes a resource intensive mission while the network size increasing.
ML can help to form the basis for network performance measurement. Moreover, by applying ML in network performance measurement, efficiency and resiliency can potentially be improved. 

In this survey article, we have comprehensively covered the advantages of ML-based network performance measurement. We have then discussed case studies on the use of network performance measurement with ML. Finally, we have identified and discussed challenges, issues, and future research directions related to ML-based network performance measurement before concluding the paper.

\bibliographystyle{unsrtnat}
\bibliography{references}  

\begin{thebibliography}{15}
\providecommand{\natexlab}[1]{#1}
\providecommand{\url}[1]{\texttt{#1}}
\expandafter\ifx\csname urlstyle\endcsname\relax
  \providecommand{\doi}[1]{doi: #1}\else
  \providecommand{\doi}{doi: \begingroup \urlstyle{rm}\Url}\fi

\bibitem[Kiefer et~al.(2010)Kiefer, Gollan, and Schmitt]{KGS10-1}
Andreas Kiefer, Nicos Gollan, and Jens~B. Schmitt.
\newblock Searching for tight performance bounds in feed-forward networks.
\newblock In Bruno Müller-Clostermann, Klaus Echtle, and Erwin~P. Rathgeb,
  editors, \emph{15th International GI/ITG Conference on "Measurement,
  Modelling and Evaluation of Computing Systems" and "Dependability and Fault
  Tolerance" (MMB/DFT 2010)}, volume 5987 of \emph{Lecture notes in Computer
  Science}, pages 227--241, Essen, Germany, March 2010. GI/ITG, Springer.
\newblock URL \url{publications/KGS10-1.pdf}.

\bibitem[Geyer(2017)]{10.1145/3150928.3150941}
Fabien Geyer.
\newblock Performance evaluation of network topologies using graph-based deep
  learning.
\newblock In \emph{Proceedings of the 11th EAI International Conference on
  Performance Evaluation Methodologies and Tools}, VALUETOOLS 2017, page
  20–27, New York, NY, USA, 2017. Association for Computing Machinery.
\newblock ISBN 9781450363464.
\newblock \doi{10.1145/3150928.3150941}.
\newblock URL \url{https://doi.org/10.1145/3150928.3150941}.

\bibitem[Geyer and Carle(2018)]{10.1145/3229607.3229610}
Fabien Geyer and Georg Carle.
\newblock Learning and generating distributed routing protocols using
  graph-based deep learning.
\newblock In \emph{Proceedings of the 2018 Workshop on Big Data Analytics and
  Machine Learning for Data Communication Networks}, Big-DAMA '18, page
  40–45, New York, NY, USA, 2018. Association for Computing Machinery.
\newblock ISBN 9781450359047.
\newblock \doi{10.1145/3229607.3229610}.
\newblock URL \url{https://doi.org/10.1145/3229607.3229610}.

\bibitem[Li et~al.(2016)Li, Zemel, Brockschmidt, and Tarlow]{li2016gated}
Yujia Li, Richard Zemel, Marc Brockschmidt, and Daniel Tarlow.
\newblock Gated graph sequence neural networks.
\newblock In \emph{Proceedings of ICLR'16}, April 2016.
\newblock URL
  \url{https://www.microsoft.com/en-us/research/publication/gated-graph-sequence-neural-networks/}.

\bibitem[{Geyer} and {Carle}(2018)]{8493033}
F.~{Geyer} and G.~{Carle}.
\newblock The case for a network calculus heuristic: Using insights from data
  for tighter bounds.
\newblock In \emph{2018 30th International Teletraffic Congress (ITC 30)},
  volume~02, pages 43--48, 2018.
\newblock \doi{10.1109/ITC30.2018.10060}.

\bibitem[{Geyer} and {Bondorf}(2020)]{9219693}
F.~{Geyer} and S.~{Bondorf}.
\newblock On the robustness of deep learning-predicted contention models for
  network calculus.
\newblock In \emph{2020 IEEE Symposium on Computers and Communications (ISCC)},
  pages 1--7, 2020.
\newblock \doi{10.1109/ISCC50000.2020.9219693}.

\bibitem[{Bannour} et~al.(2018){Bannour}, {Souihi}, and {Mellouk}]{8187644}
F.~{Bannour}, S.~{Souihi}, and A.~{Mellouk}.
\newblock Distributed sdn control: Survey, taxonomy, and challenges.
\newblock \emph{IEEE Communications Surveys Tutorials}, 20\penalty0
  (1):\penalty0 333--354, 2018.
\newblock \doi{10.1109/COMST.2017.2782482}.

\bibitem[{Gulyas} and {Biro}(2006)]{1633336}
A.~{Gulyas} and J.~{Biro}.
\newblock A stochastic extension of network calculus for workload loss
  examinations.
\newblock \emph{IEEE Communications Letters}, 10\penalty0 (5):\penalty0
  399--401, 2006.
\newblock \doi{10.1109/LCOMM.2006.1633336}.

\bibitem[{Deng} and {Lin}(2010)]{5471427}
Y.~{Deng} and C.~{Lin}.
\newblock An extended stochastic loss bound with moment generating function.
\newblock In \emph{2010 International Conference on Communications and Mobile
  Computing}, volume~1, pages 498--502, 2010.
\newblock \doi{10.1109/CMC.2010.64}.

\bibitem[{Wang} et~al.(2013){Wang}, {Schmitt}, and {Ciucu}]{6662955}
H.~{Wang}, J.~{Schmitt}, and F.~{Ciucu}.
\newblock Performance modelling and analysis of unreliable links with
  retransmissions using network calculus.
\newblock In \emph{Proceedings of the 2013 25th International Teletraffic
  Congress (ITC)}, pages 1--9, 2013.
\newblock \doi{10.1109/ITC.2013.6662955}.

\bibitem[{Scheffler} et~al.(2018){Scheffler}, {Fögen}, and {Bondorf}]{8514938}
A.~{Scheffler}, M.~{Fögen}, and S.~{Bondorf}.
\newblock The deterministic network calculus analysis: Reliability insights and
  performance improvements.
\newblock In \emph{2018 IEEE 23rd International Workshop on Computer Aided
  Modeling and Design of Communication Links and Networks (CAMAD)}, pages 1--6,
  2018.
\newblock \doi{10.1109/CAMAD.2018.8514938}.

\bibitem[{Fei Yu} and {Krishnamurthy}(2006)]{1611082}
{Fei Yu} and V.~{Krishnamurthy}.
\newblock Effective bandwidth of multimedia traffic in packet wireless cdma
  networks with lmmse receivers: a cross-layer perspective.
\newblock \emph{IEEE Transactions on Wireless Communications}, 5\penalty0
  (3):\penalty0 525--530, 2006.
\newblock \doi{10.1109/TWC.2006.1611082}.

\bibitem[{Fidler}(2010)]{5415864}
M.~{Fidler}.
\newblock Survey of deterministic and stochastic service curve models in the
  network calculus.
\newblock \emph{IEEE Communications Surveys Tutorials}, 12\penalty0
  (1):\penalty0 59--86, 2010.
\newblock \doi{10.1109/SURV.2010.020110.00019}.

\bibitem[Khangura et~al.(2019)Khangura, Fidler, and Rosenhahn]{KHANGURA201918}
Sukhpreet~Kaur Khangura, Markus Fidler, and Bodo Rosenhahn.
\newblock Machine learning for measurement-based bandwidth estimation.
\newblock \emph{Computer Communications}, 144:\penalty0 18--30, 2019.
\newblock ISSN 0140-3664.
\newblock \doi{https://doi.org/10.1016/j.comcom.2019.05.005}.
\newblock URL
  \url{https://www.sciencedirect.com/science/article/pii/S0140366419303457}.

\bibitem[Yin and Kaur(2016)]{10.1007/978-3-319-30505-9_30}
Qianwen Yin and Jasleen Kaur.
\newblock Can machine learning benefit bandwidth estimation at ultra-high
  speeds?
\newblock In Thomas Karagiannis and Xenofontas Dimitropoulos, editors,
  \emph{Passive and Active Measurement}, pages 397--411, Cham, 2016. Springer
  International Publishing.
\newblock ISBN 978-3-319-30505-9.

\end{thebibliography}






\end{document}